\documentclass[]{AO4ELT}  

\usepackage{microtype}
\usepackage[sorting=none]{biblatex}
\usepackage{amsmath,amsfonts,amssymb}
\usepackage{graphicx}
\usepackage{pst-all} 
\usepackage[colorlinks=true, allcolors=blue]{hyperref}
\makeatletter         
\def\@maketitle{
\includegraphics[width = 170mm]{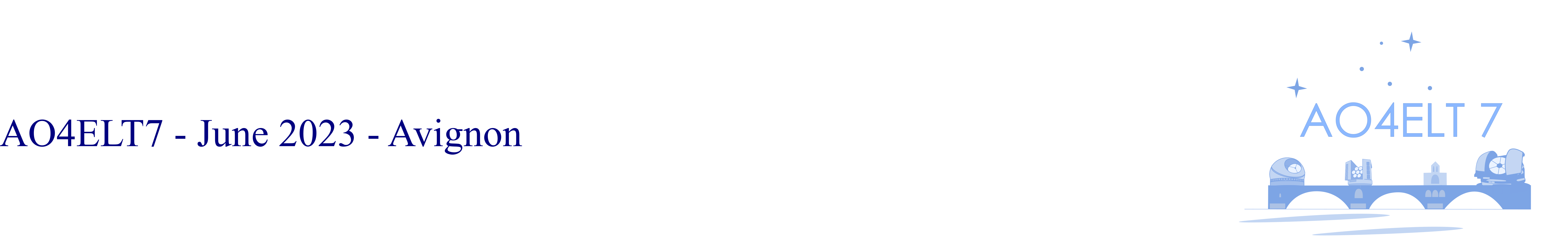}\\[8ex]
\begin{center}
{\Huge \bfseries \sffamily \@title }\\[4ex] 
{\Large  \@author}\\[4ex] 
\@date
\end{center}}

\usepackage[sc, hang]{caption}
\usepackage{graphics}
\usepackage{makeidx}
\usepackage[pass]{geometry}
\usepackage{booktabs}
\usepackage{marvosym}
\usepackage[hang]{subfigure}
\usepackage{rotating}
\usepackage{csquotes}
\usepackage{url}
\usepackage{multicol}
\usepackage{multirow}
\usepackage{overpic}
\usepackage{commath}
\usepackage{bm}
\usepackage{rotating}
\usepackage{psfrag}
\usepackage{parallel}
\usepackage{fancyhdr}

\usepackage[english]{babel} 
\usepackage{eso-pic,graphicx}
\usepackage{enumerate}
\usepackage{psfrag}
\usepackage{booktabs}

\usepackage{color, colortbl}
\definecolor{LightGray}{gray}{0.9}
\usepackage{diagbox}
\usepackage{array}
\newcolumntype{C}[1]{>{\centering\let\newline\\\arraybackslash\hspace{0pt}}m{#1}}
\usepackage{multirow}

\title{MORFEO enters final design phase}

\author[a]{Lorenzo Busoni}
\author[a]{Guido Agapito}
\author[a]{Alessandro Ballone}
\author[a]{Alfio Puglisi}
\author[b]{Alexander Goncharov}
\author[a]{Amedeo Petrella}
\author[a]{Amico Di Cianno}
\author[a]{Andrea Balestra}
\author[a]{Andrea Baruffolo}
\author[a]{Andrea Bianco}
\author[a]{Andrea Di Dato}
\author[a]{Angelo Valentini}
\author[a]{Benedetta Di Francesco}
\author[c]{Benoit Sassolas}
\author[a]{Bernardo Salasnich}
\author[a]{Carmelo Arcidiacono}
\author[a]{Cédric Plantet}
\author[a]{Christian Eredia}
\author[a]{Daniela Fantinel}
\author[a]{Danilo Selvestrel}
\author[b]{Deborah Malone}
\author[a]{Demetrio Magrin}
\author[a]{Domenico D'Auria}
\author[a]{Edoardo Redaelli}
\author[a]{Elena Carolo}
\author[a]{Elia Costa}
\author[a]{Elisa Portaluri}
\author[a]{Enrico Cascone}
\author[a]{Enrico Giro}
\author[a]{Federico Battaini}
\author[a]{Francesca Annibali}
\author[a]{Fulvio Laudisio}
\author[a]{Gabriele Rodeghiero}
\author[f]{Gabriele Umbriaco}
\author[d]{Gael Chauvin}
\author[a]{Gianluca Di Rico}
\author[a]{Giorgio Pariani}
\author[a]{Giulia Carlà}
\author[a]{Giulio Capasso}
\author[f]{Giuseppe Cosentino}
\author[d]{Jean Jacques Correia}
\author[a]{Italo Foppiani}
\author[a]{Ivan Di Antonio}
\author[a]{Jacopo Farinato}
\author[a]{Kalyan Kumar Radhakrishnan}
\author[d]{Laurence Gluck}
\author[c]{Laurent Pinard}
\author[a]{Luca Marafatto}
\author[a]{Marcello Agostino Scalera}
\author[a]{Marco Gullieuszik}
\author[a]{Marco Bonaglia}
\author[a]{Marco Riva}
\author[a]{Marco Xompero}
\author[a]{Maria Bergomi}
\author[a]{Matteo Aliverti}
\author[a]{Matteo Genoni}
\author[a]{Matteo Munari}
\author[a]{Mauro Dolci}
\author[c]{Michel Christophe}
\author[a]{Michele Cantiello}
\author[a]{Mirko Colapietro}
\author[b]{Nicholas Devaney}
\author[a]{Nicolò Azzaroli}
\author[a]{Paolo Grani}
\author[a]{Paolo Ciliegi}
\author[d]{Patrick Rabou}
\author[d]{Philippe Feautrier}
\author[a]{Pietro Schipani}
\author[a]{Roberto Ragazzoni}
\author[a]{Rosanna Sordo}
\author[a]{Runa Briguglio}
\author[a]{Salvatore Lampitelli}
\author[a]{Salvatore Savarese}
\author[a]{Simone Benedetti}
\author[a]{Simone Di Filippo}
\author[a]{Simone Esposito}
\author[a]{Simonetta Chinellato}
\author[e]{Sylvain Oberti}
\author[d]{Sylvain Rochat}
\author[a]{Tommaso Lapucci}
\author[a]{Ugo Di Giammatteo}
\author[a]{Vincenzo Cianniello}
\author[a]{Vincenzo De Caprio}
\author[d]{Zoltan Hubert}

\affil[a]{Istituto Nazionale di Astrofisica (INAF), Italy}
\affil[b]{The National University of Ireland, Galway, Ireland}
\affil[c]{Laboratoire des Matériaux Avancés (LMA) -- Institut de Physique des 2 Infinis de Lyon (IP2I Lyon), Lyon, France}
\affil[d]{Institut des Sciences de l’Univers (INSU) -- Institut de Planétologie et d’Astrophysique de Grenoble (IPAG), Grenoble, France}
\affil[e]{European Southern Observatory (ESO), Garching, Germany}
\affil[f]{Università di Bologna, Bologna, Italy}

\authorinfo{Further author information: (Send correspondence to L.B.)\\L.B.: E-mail: lorenzo.busoni@inaf.it}

\begin{document} 
\maketitle

\begin{abstract}
MORFEO (Multi-conjugate adaptive Optics Relay For ELT Observations, formerly MAORY), the MCAO system for the ELT, will provide diffraction-limited optical quality to the large field camera MICADO. MORFEO has officially passed the Preliminary Design Review and it is entering the final design phase. We present the current status of the project, with a focus on the adaptive optics system aspects and expected milestones during the next project phase.
\end{abstract}

\keywords{MORFEO, Extremely Large Telescope, European Extremely Large Telescope, Multi-Conjugate Adaptive Optics, status, preliminary design review, phase C, final design, MICADO, update}

\section{Introduction}
\label{sect:intro}  

The Multi-conjugate adaptive Optics Relay For ELT Observations (MORFEO, formerly known as MAORY)~\cite{2021Msngr.182...13C,2022SPIE12185E..14C} will deliver a multiconjugate adaptive optics (MCAO) correction to the first light instrument MICADO (Multi-AO Imaging Camera for Deep Observations)~\cite{2021Msngr.182...17D} and to a future second port instrument of the ESO Extremely Large Telescope (ELT)~\cite{2020SPIE11445E..1ET}.
The MCAO correction of MORFEO is provided by the fourth mirror of the ELT (M4), that is conjugated at 600 m, and by two post-focal deformable mirrors (DMs) conjugated at 6.5 and 17.5 km.
The DMs compensation is based on the measurement from both three natural guide star (NGS) wavefront sensors (WFSs) and six laser guide star (LGS) WFSs.
MORFEO is required to provide a uniform correction over a 1-arcmin field of view, with a Strehl Ratio of 30$\%$ (goal 50$\%$) in K band under ESO's median atmospheric conditions and with 50$\%$ sky coverage, allowing MICADO to exploit the full resolution potential of the 39-m ELT aperture in the 0.8 - 2.4 $\mu$m wavelength coverage.
In particular, MICADO will take advantage of MORFEO correction in imaging and spectroscopy observing modes. For what concerns the imaging mode, two options are foreseen: a low-resolution imager, covering a wide field of view of 50.5 $\times$ 50.5 arcsec$^2$ at a pixel scale of 4 mas, and a high-resolution imager covering a smaller field of view of 19 $\times$ 19 arcsec$^2$ at a pixel scale of 1.5 mas.
This will allow a significant step forward for several science cases, with a significant increase in spatial resolution with respect to HST and even JWST (pixel scale $\sim$ 30mas/pix).
For what concerns spectroscopy, the full wavelength range is covered with two settings: a short slit for 0.84 - 1.48 $\mu$m and a long slit 1.48 - 2.46 $\mu$m.

\begin{figure}[h]
    \centering
    \includegraphics[width=0.75\linewidth]{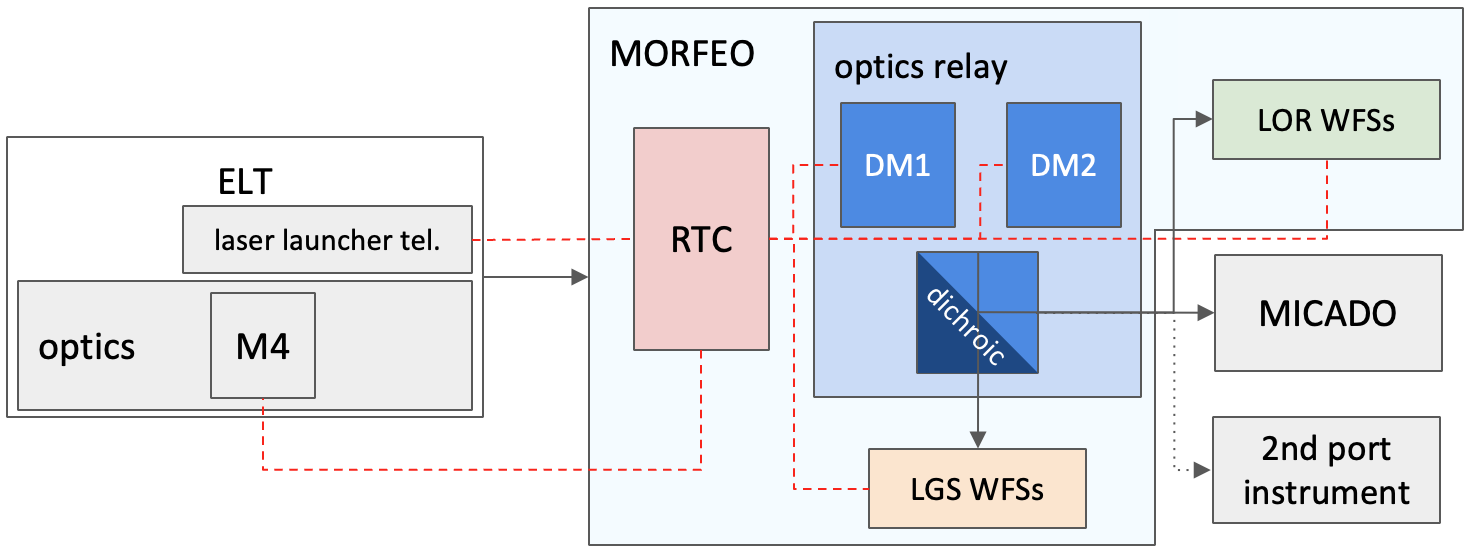}
    \caption{Diagram of the ELT, MORFEO, MICADO and the second port instrument. Light is coming from ELT which also provides the source for the 6 laser guide stars (LGS) and the ground layer correction with M4. MORFEO interacts with the telescope and, in particular, with the 6 laser launch telescopes and M4 through the real-time computer (RTC). In addition, the RTC receives the measurements from the wavefront sensors (WFSs) and commands the two deformable mirrors (DMs) of MORFEO (for more details about the WFSs, DMs and RTC see Secs. \ref{sect:WFSs}, \ref{sect:DMs} and \ref{sect:RTC} respectively). One dichroic splits the laser light (589nm) and transmits it to the LGS WFSs. Finally, the corrected wavefront is provided to MICADO and to the NGS (LOR) WFS module (accessing 3 NGS stars in an outer annulus around the MICADO field),  or to a second port thanks to a flat flip mirror, M11M (for more details about the optical design see Sec. \ref{sect:optics}).}
    \label{fig:MORFEO_scheme}
\end{figure}
\begin{figure}[h]
    \centering
    \includegraphics[width=0.85\linewidth]{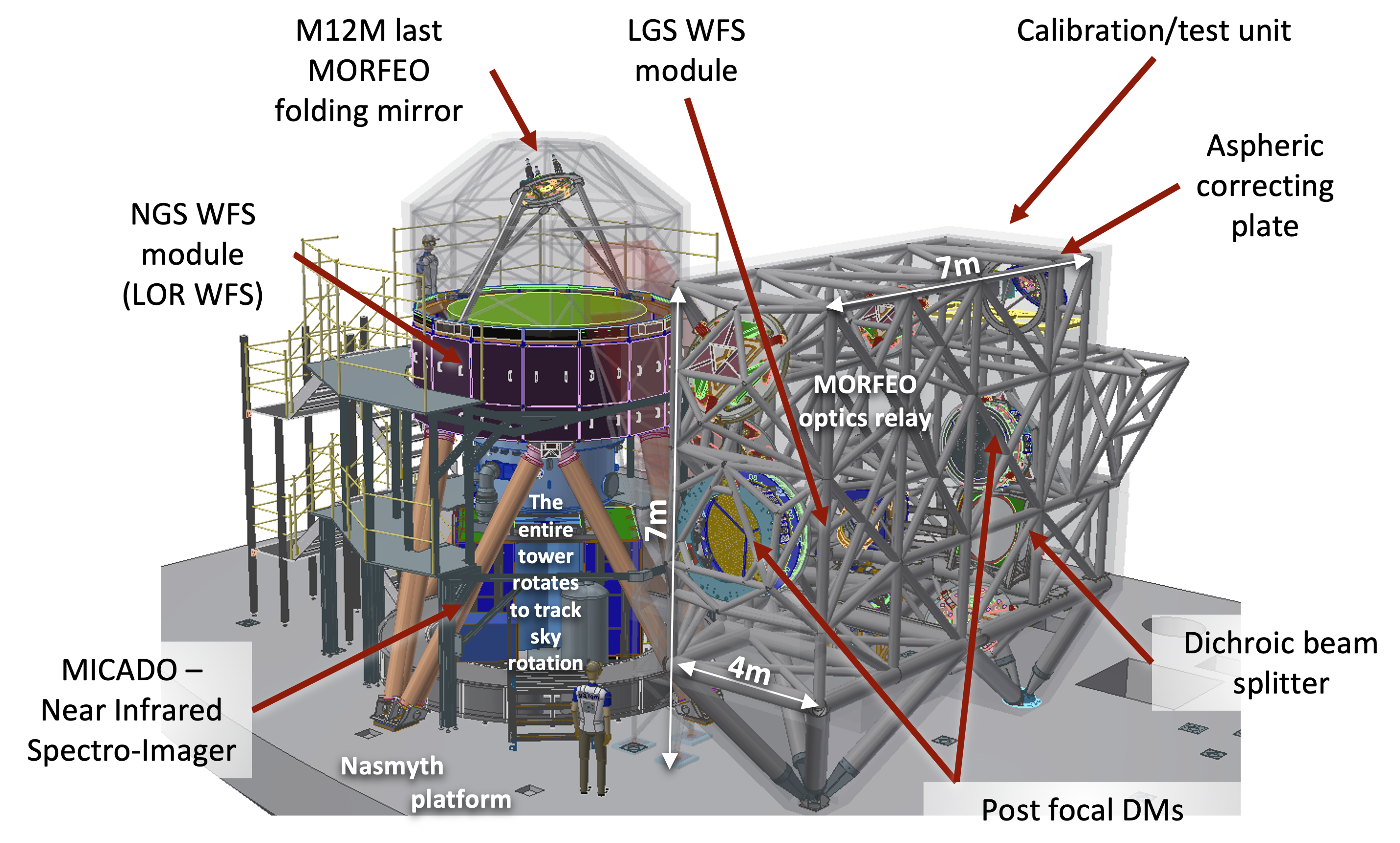}
    \caption{3D model of the ELT Nasmyth platform with MORFEO and MICADO. The main elements of MORFEO are labeled in the image.}
    \label{fig:MORFEO_3D}
\end{figure}

A diagram of the ELT, MORFEO, MICADO, and the second port instrument is shown in Fig.~\ref{fig:MORFEO_scheme}: the ELT provides the main deformable mirror of MORFEO, M4~\cite{2019Msngr.178....3V}, the laser launch telescopes and the prefocal station (PFS)~\cite{2020SPIE11445E..1GL}; the MORFEO relay provides the corrected beam to MICADO or to a future second port instrument and consists of 2 postfocal deformable mirrors, of the LGS WFS module and of the NGS WFS module mounted on the MICADO derotator; the MORFEO Real-Time Computer (RTC)~\cite{2022SPIE12185E..5KB} controls the MORFEO hardware and interacts with the telescope, with M4 for the ground layer correction and with the launcher telescopes to correct the laser jitter and pointing.
A 3D model of the ELT Nasmyth platform with MORFEO and MICADO is shown in Fig. \ref{fig:MORFEO_3D}: the MORFEO optical relay~\cite{2022SPIE12185E..5MP} is housed in a large mechanical structure of 7$\times$7$\times$4 m (see Refs. \cite{2022SPIE12184E..2XD} and \cite{2022SPIE12184E..3RC}), that will be shielded to increase thermal inertia, reduce airflow and local turbulence~\cite{2022SPIE12185E..4PA}.
At the entrance port of MORFEO, close to the ELT focal plane, is installed the calibration unit~\cite{2022SPIE12185E..5FD} that can feed the WFS modules and MICADO through the relay; the LGS WFS module~\cite{2018SPIE10703E..1YS} is not clearly visible in the picture, hidden by the structure.
The last folding mirror and the NGS WFSs~\cite{2022SPIE12185E..4OB} are above MICADO with the NGS WFSs that rotate together with MICADO to follow the sky rotation. The second port instrument will have to implement its own NGS WFS module.

The MORFEO project successfully closed the Preliminary Design Review (PDR) with ESO in February 2023.  During the current phase, we will finalize the design that will be 
reviewed with ESO during the Final Design Review (FDR) foreseen by 2025. For several subsystems the final design will be developed in collaboration with external companies, which, after the successful conclusion of the FDR, will also be in charge of the manufacturing. Successively, the assembly process will be carried out at the Integration hall located at the INAF premises in Bologna.


\section{Science}
\label{sect:science}  

The development and implementation of the science cases are managed for the MORFEO consortium by the MORFEO Science Team, which works in close collaboration with the MICADO Science Team and operates to maximize the scientific return from the ELT-MORFEO-MICADO system.
The MORFEO science team has widely explored the science cases for the MORFEO+MICADO.   A preliminary collection of the cases studied is reported in the MORFEO science cases white book freely available on the MORFEO website \cite{MAORYScienceCasesWhiteBook}.

The collected science cases address many of the major questions of astrophysics: 

\begin{itemize}
    \item Planetary systems, including cases in our own Solar Systems, exoplanets, and formation of planetary systems. 
    \item Nearby stellar systems with stars and stellar systems within our own Galaxy and its satellite.
    \item Local Universe with science cases aimed to study the stellar content and the structure of distant stellar systems that can be at least partially resolved into individual stars. In many cases, they will fall within the range of resolved systems only thanks to the advent of MORFEO+MICADO.
    \item High-redshift Universe with the science cases addressing the formation of structures and cosmology using the formidable sensitivity and resolution of MORFEO+MICADO to probe the very distant Universe and, consequently, the earliest phases of galaxy formation, as well as high energy phenomena over the range of cosmic distance and times made accessible by ELT.   
\end{itemize}

\section{Optical design}
\label{sect:optics}  

\begin{figure}[h]
    \centering
    \includegraphics[width=\linewidth]{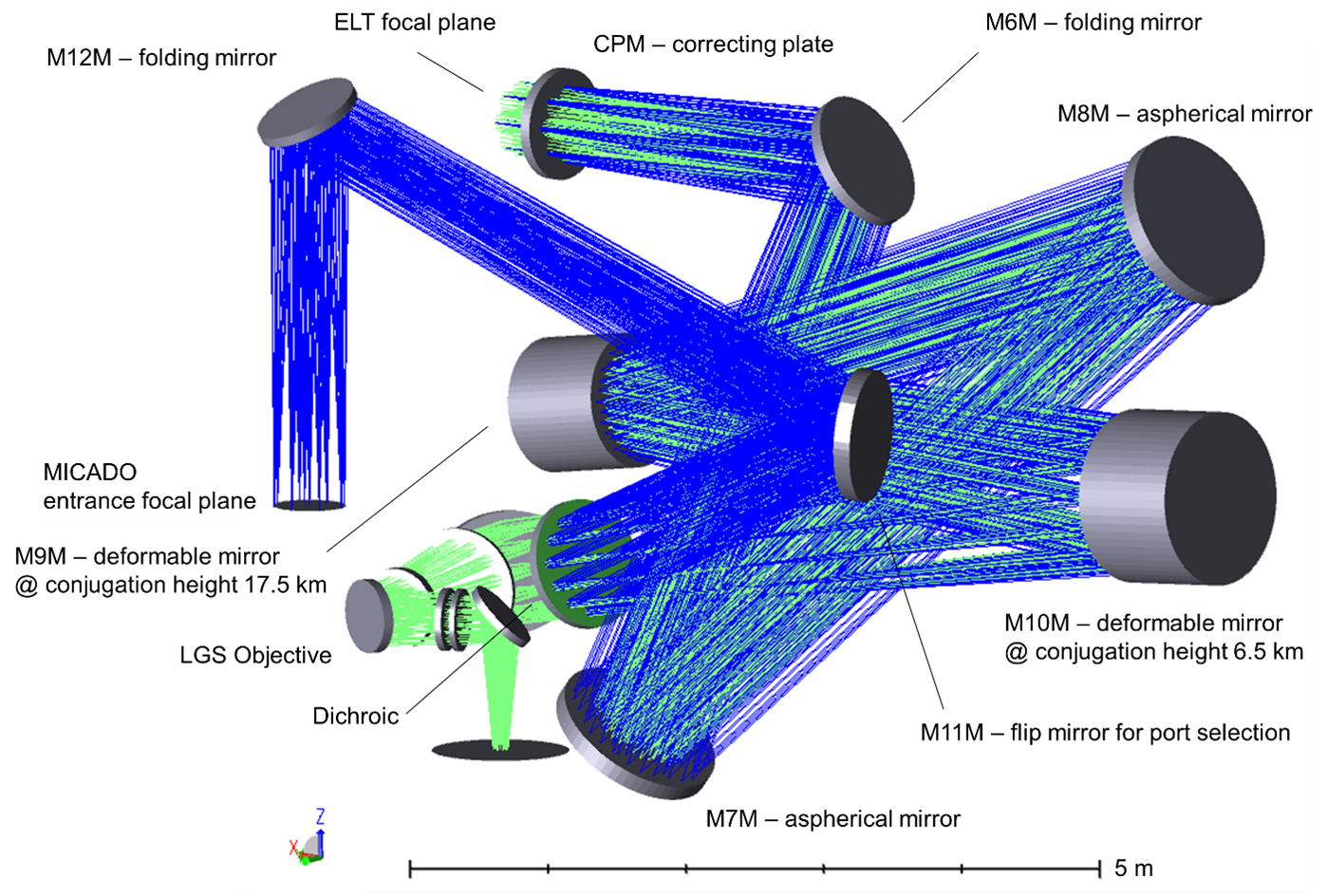}
    \caption{3D view of the MORFEO optical configuration. Blue rays represent the science and NGS path to MICADO, green rays represent the LGS path.}
    \label{fig:OPTICS}
\end{figure}

The present configuration has been initially proposed by Bernard Delabre, and, in a similar arrangement, by Johan Kosmalski. It is based on four powered mirrors (two aspherical and two spherical) able to deliver high quality in terms of WFE and field distortion. It is tolerant of defocus and astigmatism that can be compensated for by acting on the last mirror with power. This configuration implements the concept, firstly introduced by Andrew Rakich, of using an aspherical plate near the telescope focal surface to correct multiple object conjugate distances (see Refs. \cite{2020SPIE11451E..4JR}, \cite{2020SPIE11548E..0MR} and \cite{2022SPIE12185E..0CR}).
The correcting plate (CPM) greatly improves the quality of the exit pupil image, of the DMs meta-pupil images and of the LGS images at different conjugation altitudes without degrading the quality for the infinite conjugated surface. 
We have adapted and re-optimized the general optical solution to the MORFEO specific requirements and constraints, namely, the available volume on the Nasmyth platform and the position of the MICADO entrance surface. Folding mirrors were added to place the instrument in a vertical layout and to provide switching between MICADO and the second instrument. A scheme is shown in Fig.~\ref{fig:OPTICS} (see Refs. \cite{2020SPIE11448E..34M} and \cite{2022SPIE12185E..5MP}).

\subsection{Main Path Optics description}
The F/17.75 beam coming from the ELT focal surface initially passes through the CPM. 
Following the path, there is a first flat mirror, M6M, folding the beam down to the first aspherical concave mirror M7M. The beam is reflected up to the second aspherical concave mirror M8M, and then down again towards the first DM, M9M, having a spherical surface. This mirror is the only convex surface on the main path. After that, the beam is reflected by the second DM, M10M, which is concave and spherical. After M10M, the image of the pupil is formed, and just after this image, the science and NGS light is separated from the LGS light by means of a dichroic filter. The LGS light (589 nm) is transmitted, while science and NGS light (600 nm - 2400 nm) are reflected. The reflected light then reaches M11M, a flat flip mirror that allows the selection of the MICADO path or of the second instrument path through a rotational axis. Finally, the light is reflected by a flat mirror, M12M (or M12M bis), installed over MICADO (or, as previously mentioned, some future second instrument) and comes to a focus at the gravity invariant entrance focal surface of the instruments. 

\subsection{Laser Guide Star Objective description}
\label{sect:LGSO}
The LGS Objective (LGSO) reduces the LGS beam F/number at wavelength 589 nm to F/5 at the focal surface of the entrance of the LGS WFS module. It is made up of four silica spherical lenses and three-fold mirrors. The first two mirrors have been introduced in order to maintain the objective within the available envelope, while the third mirror allows to feed the LGS module with a gravity invariant and telecentric focus. The LGS asterism required by MORFEO is fixed at a field angle of 45 arcsec. As the ELT elevation varies (Zenith angle in the 0-60 degrees range), the LGS launchers will maintain the LGS sources on the sodium layer at a fixed angle of 45 arcsec. Thanks to the telecentric beam delivered by the LGSO, the radial coordinate of the LGS images on the LGSO focal surface will not change as the apparent altitude (84-180 km) of the sodium layer varies. In this way, as the telescope elevation changes, the LGS module just needs to rotate around its longitudinal axis to keep the pupil orientation and move up and down to track the sodium layer.

\section{Wavefront sensors}
\label{sect:WFSs}  

The system makes use of the 6 laser guide stars provided by the ELT and of 3 natural guide stars for low order modes correction and for truth sensing. The corresponding wavefront sensors are grouped in a LGS WFS module and in a Low Order and Reference (LOR) WFS unit, field-derotated by MICADO itself.

\subsection{The Laser Guide Star Wavefront Sensors}
\label{sect:lgs}

\begin{figure}[h]
    \centering
    \includegraphics[width=0.65\linewidth]{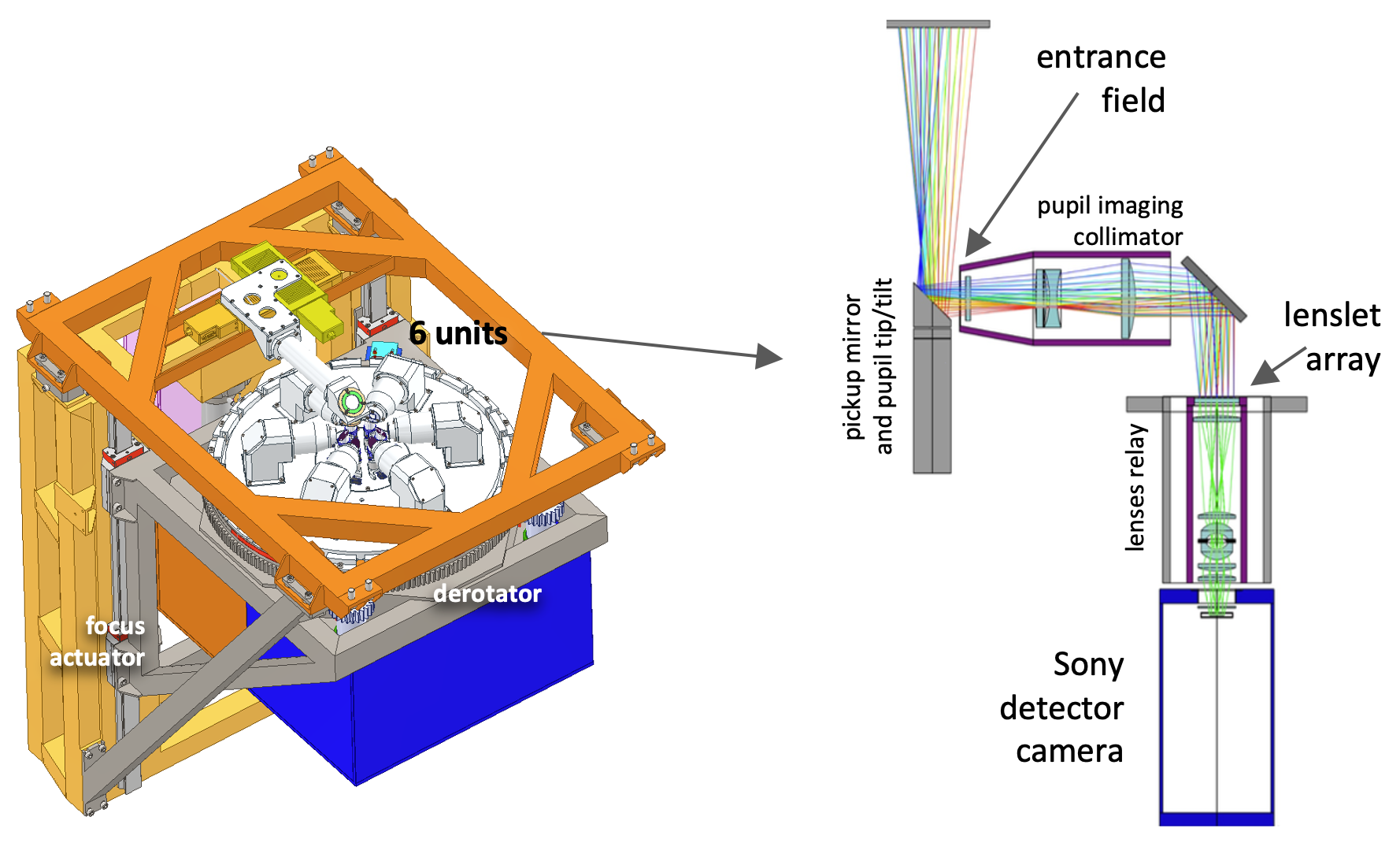}
    \caption{Laser guide star wavefront sensor (LGS WFS). Left, 3D model of the assembly of the 6 units in the mechanical structure with derotator and focus actuator, right, schematic of a single unit showing the optics and detector.}
    \label{fig:LGSWFS}
\end{figure}

The LGS WFS module is gravity invariant and it can be divided in three main units: the wave-front sensor probes, that pick off the MORFEO beam after the LGSO (see Sec. \ref{sect:LGSO}) in an LGS focal plane and sample the ELT pupil to measure the wavefront aberrations (there are 6 such probes for 6 laser launch telescopes, see right part of Fig. \ref{fig:LGSWFS}), the support structure that holds the 6 probes and aligns them in the MORFEO beam (left part of Fig. \ref{fig:LGSWFS}) and the module control electronics, cabling, power and cooling lines.

Each WFS has a 68 $\times$ 68 lenslet array positioned after a pupil imaging collimator and a folding mirror and, then, an optical relay re-images the LGS spots onto the camera detector.
As detector chip we selected the Sony IMX 425 and one of the key points that encouraged us to make this choice is the clear advantage of global shutter detector for our specific application (see Refs. \cite{2022JATIS...8b1505A} and \cite{2022SPIE12185E..8DA}).
The pick-off mirror in each WFS is mounted on a 2-axis piezo stack and allows to recenter the pupil image onto the lenslet array to compensate for possible shifts of the pupil. 
The WFS design is similar to the one of HARMONI~\cite{CostilleAO4ELT7} and provides a large FoV of about 16 arcsec on 14$\times$14 pixels per sub-aperture to minimize the truncation effect due to sodium layer extension (see Ref. \cite{2022JATIS...8b1514F}).

The support structure has two actuators: a linear stage to compensate defocus due to the varying distance of the sodium layer (it depends on telescope elevation and sodium layer average altitude) and a de-rotator to track telescope elevation and maintain a fixed orientation with respect to the telescope pupil.
More details on this module can be found in Ref.~\cite{2018SPIE10703E..1YS}.

\subsection{The Low-Order and Reference Wavefront Sensors}
\label{sect:lor}

\begin{figure}[h]
    \centering
    \includegraphics[width=0.85\linewidth]{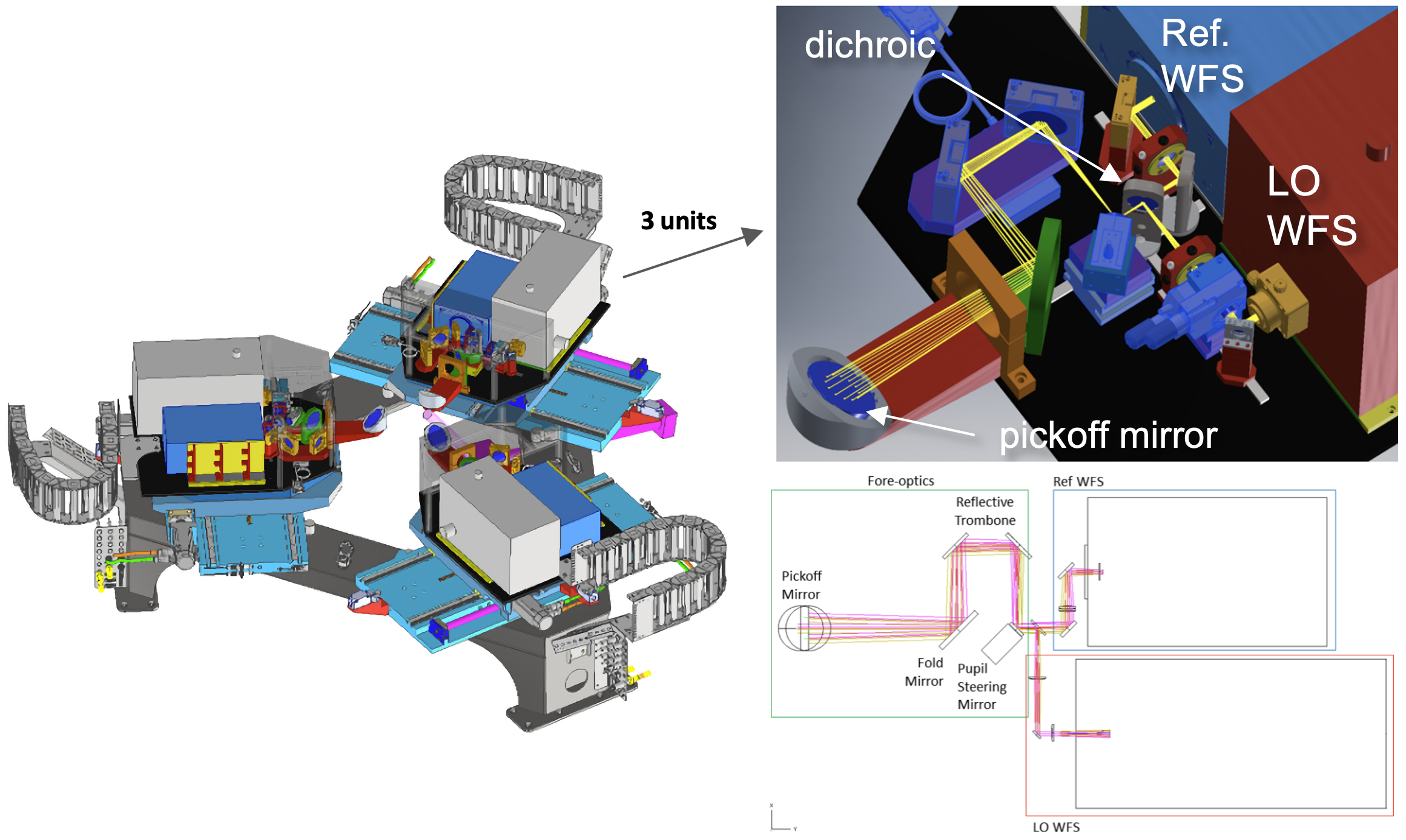}
    \caption{Natural guide star wavefront sensor (NGS WFS) known as LOR (Low Order and Reference) WFS. Left: 3D model of the assembly of the 3 units in the mechanical structure with the x and y stages; top right: schematic of a single unit showing the optics and detectors; bottom right: optical design of the LO and Ref WFS.}
    \label{fig:NGSWFS}
\end{figure}

MORFEO implements 3 NGS WFSs dedicated to: 1) sense the atmospheric low-order modes (tip, tilt and focus), 2) detrend the mid-high order modes sensed by the LGS WFS and affected by errors due to non-common path aberrations on the LGS path and sodium layer instabilities.
These 2 types of WFS take place on top of the same optical breadboard and share the light of the same NGS (infrared and visible part respectively), thus their assembly is often called Low-Order and Reference WFS~\cite{Bonaglia2022SPIE}.
The design of the first WFS, the low-order one, considers the usage of the FREDA camera~\cite{2018SPIE10703E..1WD} while the design of the Reference WFS (R WFS) makes use of ALICE~\cite{10.1117/12.2629931}.
The main design parameters of the two types of WFS are summarized in Table \ref{tab:lor_params}.
Each unit has a trombone for focus compensation and a piezo gimbal mirror for pupil recentering, as in the LGS WFSs. The Low Order arm, sensing in H band, implements an ADC. 

The three units LOR WFSs are located within the green volume on top of the MICADO assembly, as shown in Fig.~\ref{fig:MORFEO_3D}. The three LOR WFS are identical opto-mechanical assemblies, arranged on a triangular geometry and installed on top of a dedicated support structure (dark grey color on the left in Fig.~\ref{fig:NGSWFS}). Each LOR WFS is moved in a planar region of about 300 $\times$ 600 mm by a couple of orthogonal stages, called Acquisition Stages. Indeed these devices allow the LOR WFS to move across the ``patrol field'' and to reach the position where the light of the guide stars is imaged by the telescope. The whole opto-mechanical assembly of the LOR WFS is rigidly connected, and supported, by the MICADO rotator that allows for field de-rotation. The LOR WFS also implements a dedicated control electronics that is hosted in the co-rotating cabinets below the MICADO dewar. Details of the LOR WFS electronics design are given in Ref.~\cite{Lapucci_2022SPIE12185E..4NL}.

\begin{table}[h]
    \centering
    \begin{tabular}{|l|c|c|}
         \hline
         \textbf{Parameter} & \textbf{LO WFS} & \textbf{Ref WFS} \\
         \hline
         Number of WFS & 3 & 3 \\
         Wavelength band & H & R+I \\  
         Subapertures on pupil diameter & 2 & 8 \\
         Pixels on subaperture side & 128 & 30 \\
         Pixel scale [asec/pix] & 16 & 165 \\
         Field-of-view side [arcsec] & 2.048 & 4.950 \\
         \hline
    \end{tabular}
    \caption{Main parameters of the LO and Ref WFS design.}
    \label{tab:lor_params}
\end{table}

\section{Deformable mirrors}
\label{sect:DMs}  

\begin{figure}
    \begin{center}
        \subfigure[DM1 - M9M.\label{fig:dm9map}]
        {\includegraphics[width=0.4\columnwidth]{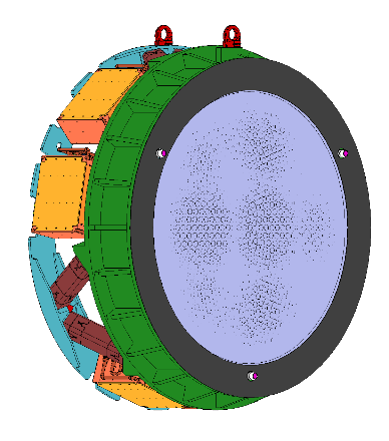}}
        \subfigure[DM2 - M10M.\label{fig:dm10map}]
        {\includegraphics[width=0.4\columnwidth]{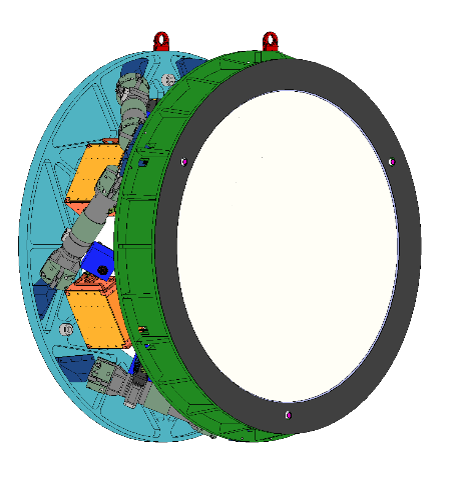}}
    \end{center}
    \caption{\label{fig:dmactmap}3D model of M9M and M10M.}
\end{figure}
MORFEO will be equipped with two post-focal deformable mirrors (DM). The current study implements the use of two  large diameters DMs (928mm for M9M and 1225 mm for M10M)  and respectively 918 and 1027 degrees of freedoms available for correction (see Fig \ref{fig:dmactmap}). In particular, according to current technical requirements M10M will be the largest adaptive mirror with monolithic glass ever built.

They will provide fast turbulence correction (baseline 500Hz) and they will be able to achieve a correction better than 40nm RMS during bad seeing conditions (1.5 arcsec at 0.5 $\mu m$ wavelength), with high reliability and minimal shape degradation up to 10\% failing actuators.

See Ref.~\cite{2022SPIE12185E..7SX} for a detailed description of the deformable mirror design.

\section{Real-time computer and Control}
\label{sect:RTC}  

\begin{figure}[h]
    \centering
    \includegraphics[width=0.65\linewidth]{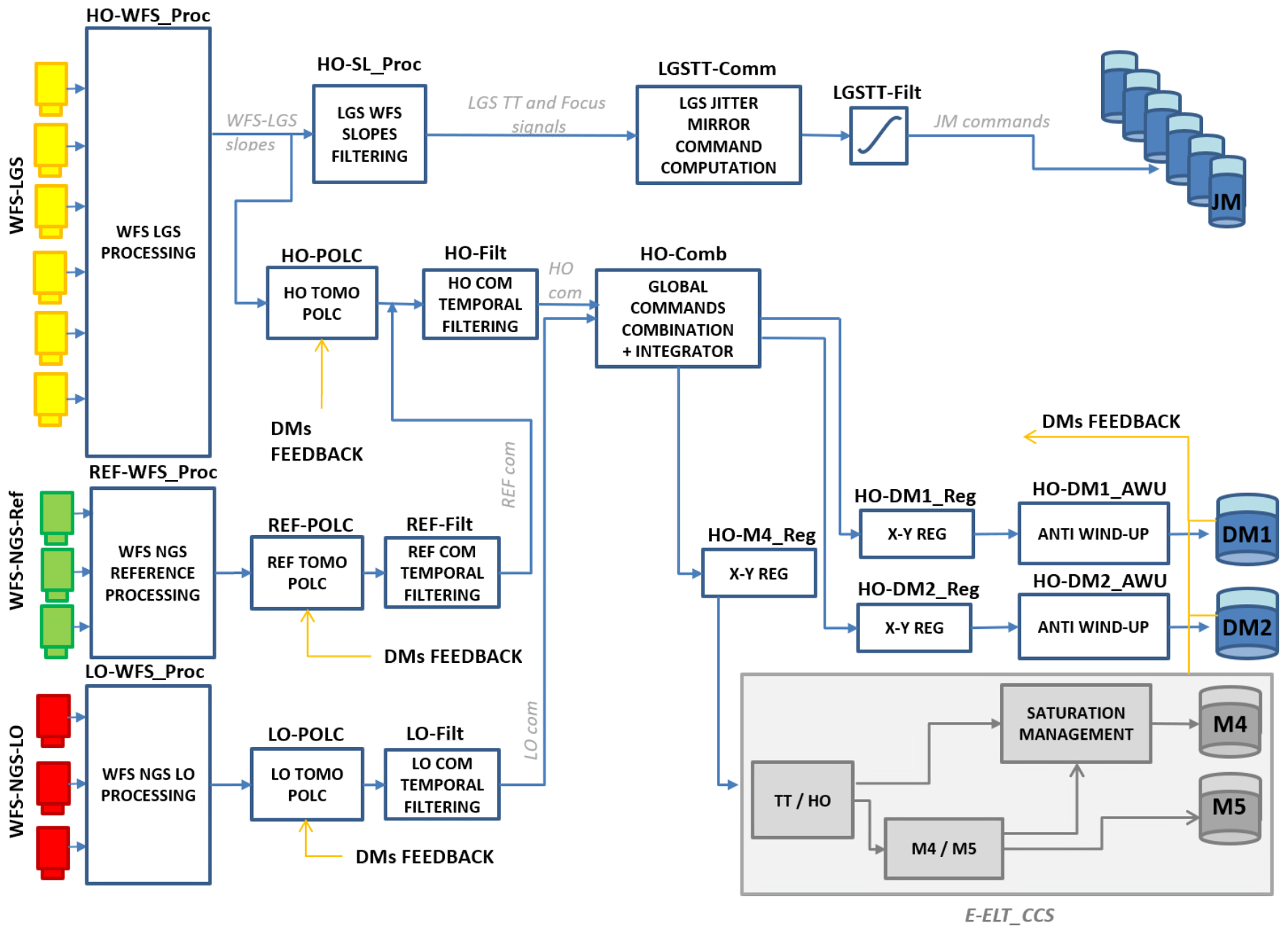}
    \caption{Real-time computer schematic.}
    \label{fig:RTC-control}
\end{figure}

MORFEO RTC is based on HEART~\cite{DunnAO4ELT7} for the hard real-time part and the ESO toolkit for the soft real-time part.
A diagram of the hard RTC part is shown in Fig. \ref{fig:RTC-control}: it receives the measurements of the 12 WFSs, computes the slopes, runs three tomographic reconstructions in parallel, apply temporal filters to the differential commands, combines the LGS and NGS commands, applies anti-wind up to avoid saturation of the DMs and computes the commands to the tip/tilt jitter actuators of the 6 laser launch telescopes.
The Control approach that will be implemented by the RTC is described in Busoni \textit{et al.}~\cite{Busoni2019} and its key elements are:
\begin{itemize}
    \item Tomography is split between high order modes and low order modes~\cite{2008JOSAA..25.2427G}.
    \item Reconstruction is done through the Minimum Mean Square Error (MMSE) estimator (alternative approaches are evaluated, as for example the one presented in Ref. \cite{StadlerAO4ELT7}).
    \item Noise priors, see Refs. \cite{2010JOSAA..27A...1B} and \cite{Oberti2019}.
    \item Pseudo-open loop control (POLC) ~\cite{2003SPIE.5169..206E}.
    \item Super-resolution, see Refs. \cite{2022A&A...667A..48O} and \cite{2022JATIS...8b1514F}.
    \item Tomographic truth sensing, see Refs. \cite{Busoni2019} and \cite{2022SPIE12185E..4RB}.
    \item Adaptive Vibration Cancellation, see Ref. \cite{7320616}.
\end{itemize}
Note that the testing activity of the MORFEO RTC tools and algorithms has already begun as shown in Refs. \cite{2023RNAAS...7....8J} and \cite{BelloneAO4ELT7}.

\section{Operation}
\label{sect:operation}  
The control strategy of MORFEO deals with the main closed loop operations, that include slope computation, wavefront reconstruction and control, and with auxiliary loops that update the system depending on external conditions variation in order to provide a stable correction during the observations.
All this is managed by the Instrument Control System Software (ICSS)~\cite{2022SPIE12189E..1VS}.
ICSS will control almost 100 functions and will interface with RTC and two different instruments (MICADO and the future second port instrument).

The main loop algorithms are implemented in the RTC, as described in Sec.~\ref{sect:RTC}, and are driven by the WFSs, whose signals are used for a tomographic reconstruction that follows a split tomography approach \cite{2022SPIE12185E..4RB}. In particular, Low Order (LO) and High Order (HO) modes have different and independent tomographic reconstructors and projection matrices: on the one hand, tip, tilt and platescale modes are reconstructed from NGS WFSs measurements and the correction foresees to command tip and tilt on M4, and platescale modes through a proper combination of focus and astigmatism on both M4 and a high-layer DM; on the other hand, the higher order modes are reconstructed from the LGS WFSs measurements. 

Standard on sky operations will be guided by MICADO and MORFEO will provide the AO correction: all MICADO template scripts will be executed by the sequencer application
that is part of the MICADO ICSS and control of AO functions from MICADO will be enabled through a dedicated API library  that encapsulates the necessary tasks within a set of Python procedures.
MORFEO will start its operation to be ready for the guide stars acquisition and for closed loop operations while the telescope is slewing to the desired target.
The operation first foresees the closure of the LGS loop to compensate for HO modes (correction of focus will be biased due to the unknown sodium layer altitude, as well as the correction of few low order modes due to truncation effects), then the closure of the NGS loop that is based on the NGS acquisition sequence described in detail in Ref.~\cite{AgapitoAO4ELT7}: the brightest NGS is acquired on the R WFS (having a larger field of view than the LO WFS) that measures tip, tilt and focus to center the star on the LO WFS and also to compensate for the focus bias; the star is then acquired on the LO WFS and tip, tilt and focus loops are closed; finally the other two NGSs are acquired in parallel to minimize the overall acquisition time.

For what concerns the auxiliary loops of MORFEO, they are summarized in the following list:
\begin{itemize}
    \item Recurrent task to update the HO loop control matrices, driven by zenith angle.
    \item Recurrent task to update the LO loop control matrices, driven by rotation angle and LOR position.
    \item Recurrent task to update the number of controlled modes, driven by seeing monitor and DMs force saturation.
    \item Recurrent task to update slope computer parameters, driven by changes in the sodium profile and on the LOR correction level.
    \item Recurrent task to update vibration rejection filters.
    \item Full-framerate loop to compensate for LGS jitter, driven by LGS WFSs tip-tilt measurements and acting on laser launcher telescopes fast steering mirrors.
    \item Offload from laser launcher telescopes fast steering mirrors to laser launcher telescopes field selecting mirrors.
    \item Truth sensor loop to detrend LGS-induced bias (truncation and NCPAs) driven by R WFS and acting on modal offset.
    \item Loop to derotate the elevation from the LGS WFSs pupil, driven by the zenith angle and acting on the LGS derotator.
    \item Loop to keep the pupil footprint stable on the lenslet array, driven by subaperture illumination pattern and acting on the pupil tilt actuator in each WFS.
    \item Loop to maintain a matching between DM actuators and WFS subapertures driven by WFS measurements and acting on M4 and on post-focal DMs projection matrices.
    \item Recurrent task to sense derotation errors driven by LO measurements and acting on MICADO derotator.
\end{itemize}

\section{Calibration and test}
\label{sect:CalibTest}  

MORFEO is a complex systems that requires a wide range of calibrations to function as intended, and several test procedures will be performed to verify its functionality and performance.
The calibrations range from the data required by the detectors, such as the conversion gain factor and bias level per readout channel, to the adjustment of geometric parameters, such as the relative displacement between the sub-aperture array and the DM actuator grid, and also include the estimation of field distortion and non-common aberrations between WFS and science path.
MORFEO is thus equipped with a calibration plan that concatenates all the calibrations tasks that are required for MORFEO to perform and achieve its specifications. 
A significant fraction of the calibration templates aims at monitoring and trending the system status and performance (e.g. throughput, alignment quality), but the MORFEO Calibration Plan does not include templates needed to verify the system performance.
Instead test activity is included in the Assembly, Integration and Test (AIT) in Europe~\cite{2022SPIE12185E..71F} and Assembly, Integration and Verification (AIV) and Commissioning in Chile.
This activity involves a set of tests to verify functionality and performance of the sub-systems and will culminate in performance tests on sky. 

\begin{figure}[h]
    \centering
    \includegraphics[width=0.85\linewidth]{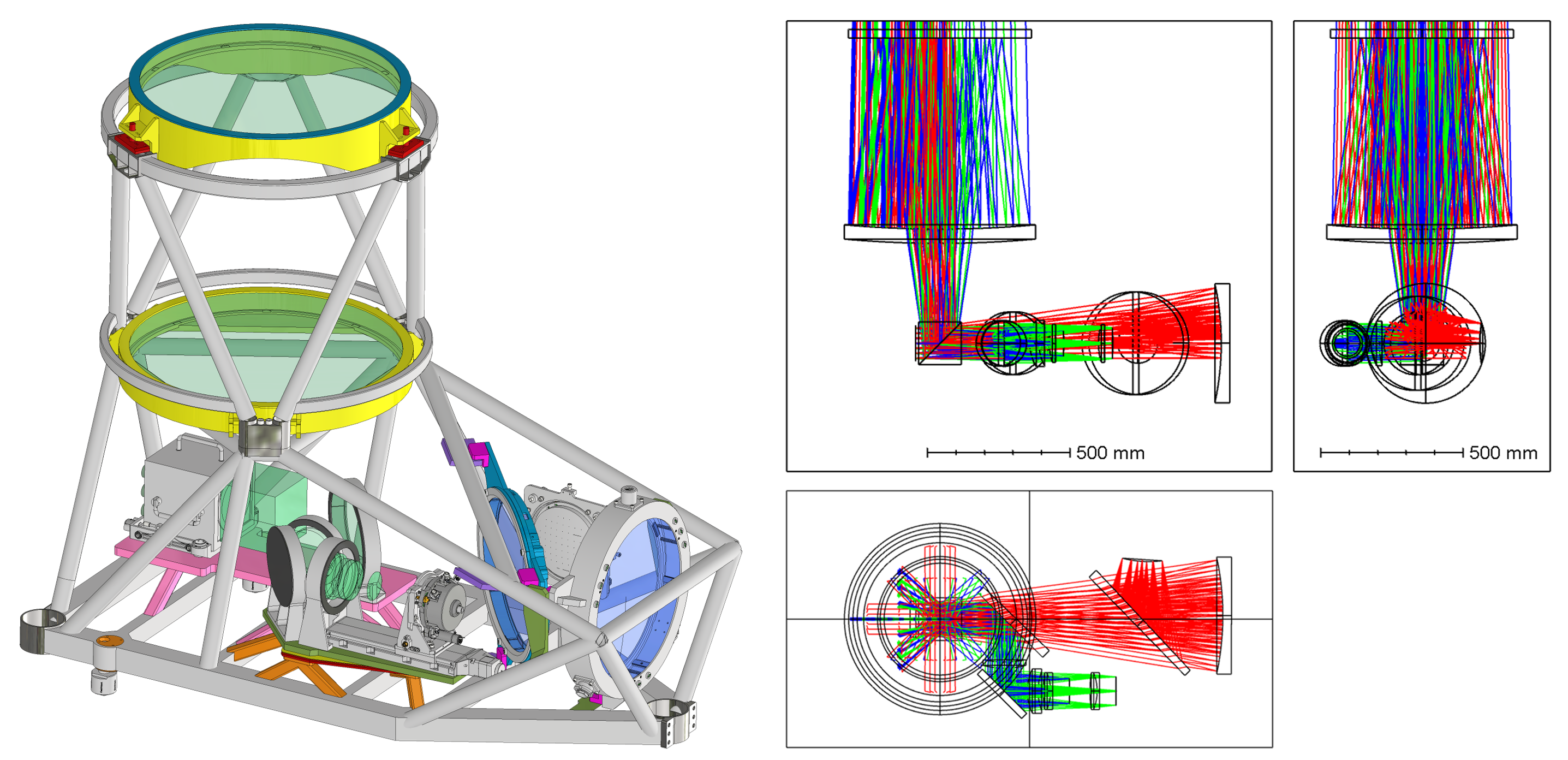}
    \caption{Calibration and Test Unit (CU and TU). On the left is a 3D model of the subsystem, on the right different views of its optical design (the folding mirror on top is not shown).}
    \label{fig:CU-TU}
\end{figure}

A relevant part of calibration and test activity will be carried out in controlled environment in the Integration Hall in Bologna, and afterwards at the telescope, thanks to a subsystem that can act both as a Calibration Unit (CU) in Chile, or as Test Unit (TU) in Italy.
The CU, shown in Fig. \ref{fig:CU-TU}, provides a set of suitable light sources, both NGS and LGS, matching the ELT optical interfaces (focal ratio, exit pupil position, field curvature), to fully simulate the telescope performance and residual aberrations. It will drastically reduce the amount of required night-time for daily, periodic and maintenance operations. A description of the current design of the CU can be found in Refs. \cite{DiRicoAO4ELT7} and \cite{DiAntonioAO4ELT7}.
The configuration of the CU can be changed to operate as TU, by installing a deformable mirror at its intermediate pupil, providing the capability to operate the full MORFEO system in standalone mode with three deformable mirrors.

\section{AO performance}
\label{sect:AOperformance}  

\begin{figure}
    \begin{center}
        \subfigure[J band, 1250nm.
        \label{fig:J_SR_Qprof_e180nm}]
        {\includegraphics[width=0.45\columnwidth]{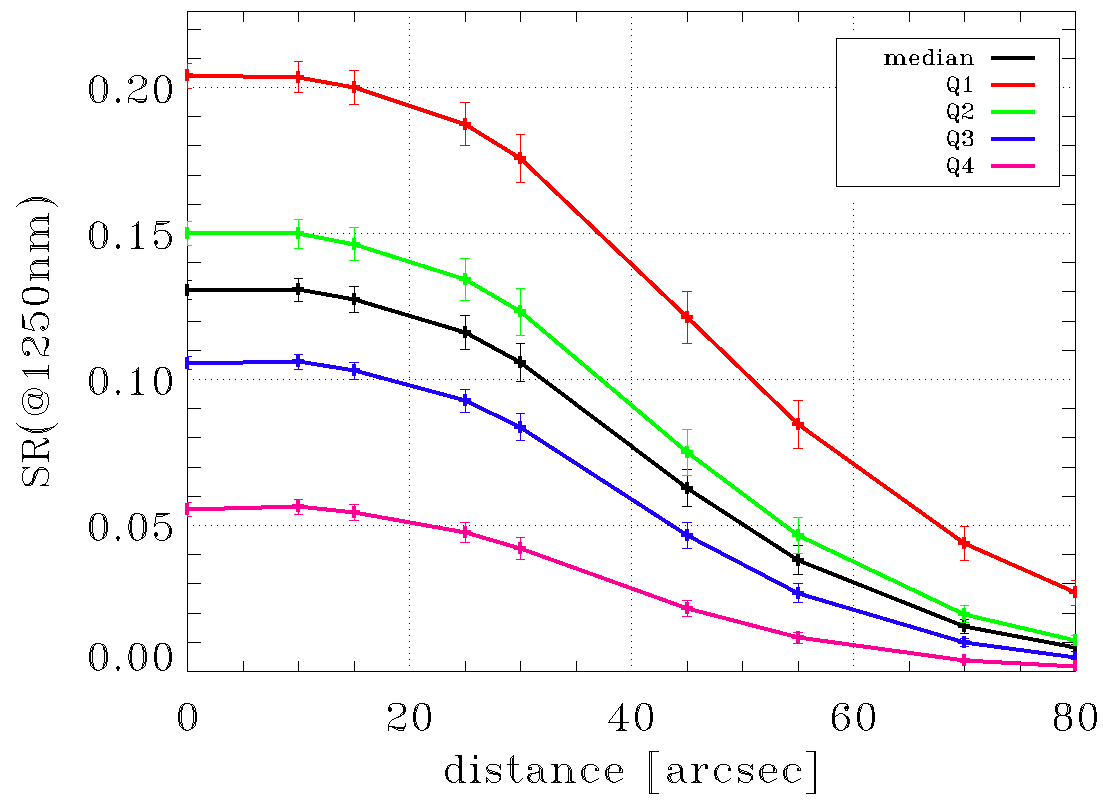}}
        \subfigure[K band, 2200nm.
        \label{fig:K_SR_Qprof_e180nm}]
        {\includegraphics[width=0.45\columnwidth]{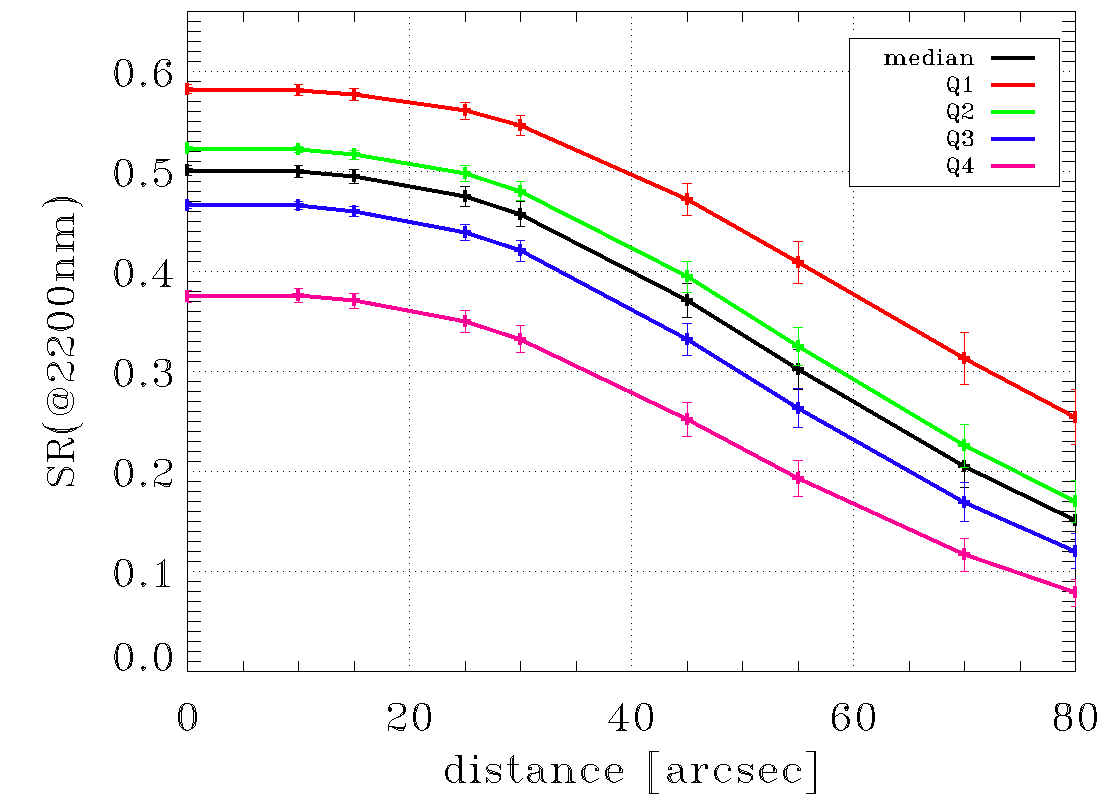}}
    \end{center}
    \caption{\label{fig:SR_Qprof_e180nm}Average SR as a function of the off-axis angle considering 5 2s-long simulations considering random realizations of each of the 5 reference atmospheric profiles considered. Zenith angle is 30deg. NGSs are bright, H=10, and at a distance of 55 arcsec. SR is estimated from PSFs in which 180nm RMS of error (from terms not considered in the simulations, see Tab. \ref{Tab:error}) are added with random phase screens with a $f^{-2}$ power spectral density. The vertical bars show the standard deviation of the SRs on the 5 random realizations. Small and large MICADO FoV is r$\le$10 arcsec and r$\le$30 arcsec respectively.}
\end{figure}
%
%
\begin{figure}
    \begin{center}
        \subfigure[J band, 1250nm, small MICADO FoV (r$\le$10 arcsec).
        \label{fig:sky_J_SR_Qprof_e180nm}]
        {\includegraphics[width=0.49\columnwidth]{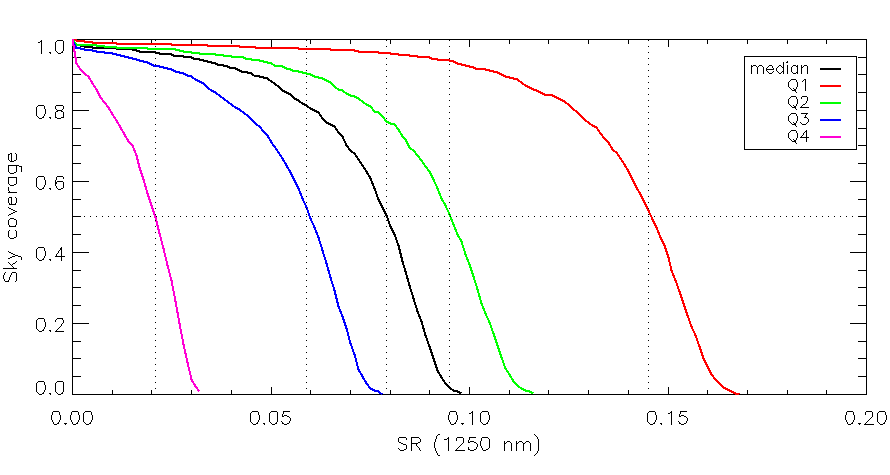}}
        \subfigure[K band, 2200nm, large MICADO FoV (r$\le$30 arcsec).
        \label{fig:sky_K_SR_Qprof_e180nm}]
        {\includegraphics[width=0.49\columnwidth]{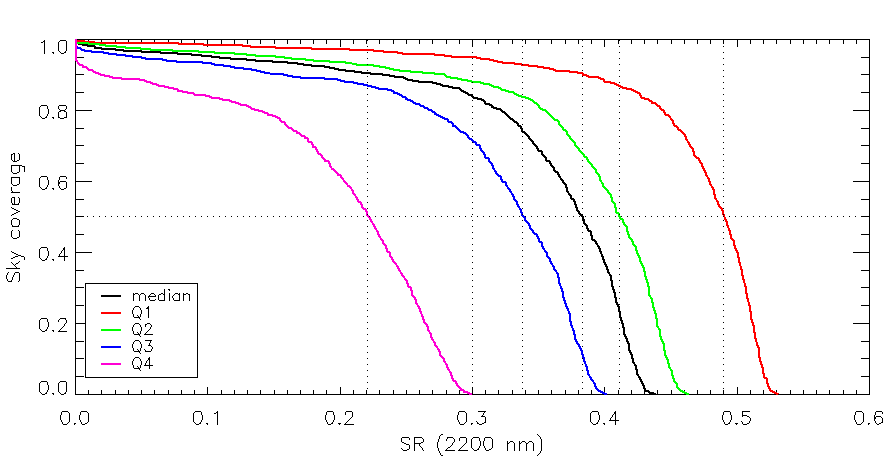}}
    \end{center}
    \caption{\label{fig:comp_sky_cov_passata_baseline2020_2PFDM_budget180nm_K_SR}Average K-band (2200 nm) SR as a function of the sky coverage at the south galactic pole for the 5 reference atmospheric profiles. Here 180nm RMS of error comes from terms not considered in the simulations (see Tab. \ref{Tab:error}). Zenith angle is 30deg.}
\end{figure}
%
%
\begin{figure}
    \begin{center}
        \subfigure[J band, 1250nm.
        \label{fig:J_FWHM_Qprof_e180nm}]
        {\includegraphics[width=0.49\columnwidth]{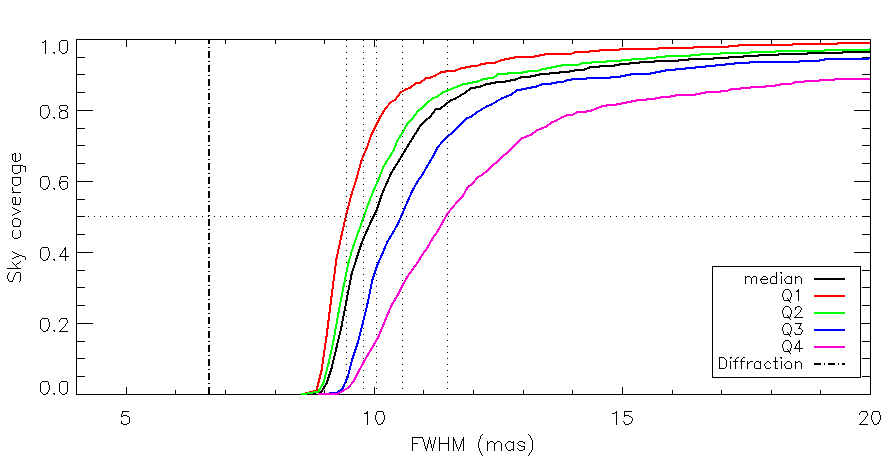}}
        \subfigure[K band, 2200nm.
        \label{fig:K_FWHM_Qprof_e180nm}]
        {\includegraphics[width=0.49\columnwidth]{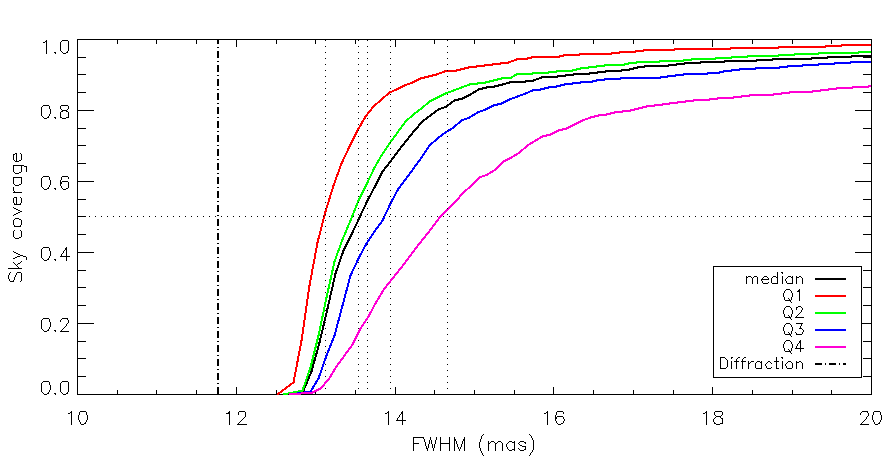}}
    \end{center}
    \caption{\label{fig:comp_sky_cov_passata_baseline2020_2PFDM_budget180nm_JHK_fwhm}Average full width at half maximum (FWHM) in the MICADO large FoV (r$\le$30 arcsec) as a function of the sky coverage at the south galactic pole considering the median atmospheric profile. The high order contribution to FWHM is estimated from PSFs in which 180nm RMS of error (from terms not considered in the simulations, see Tab. \ref{Tab:error}) are added with random phase screens with a $f^{-2}$ power spectral density. Zenith angle is 30deg.}
\end{figure}
\begin{figure}[h]
    \centering
    \includegraphics[width=0.65\linewidth]{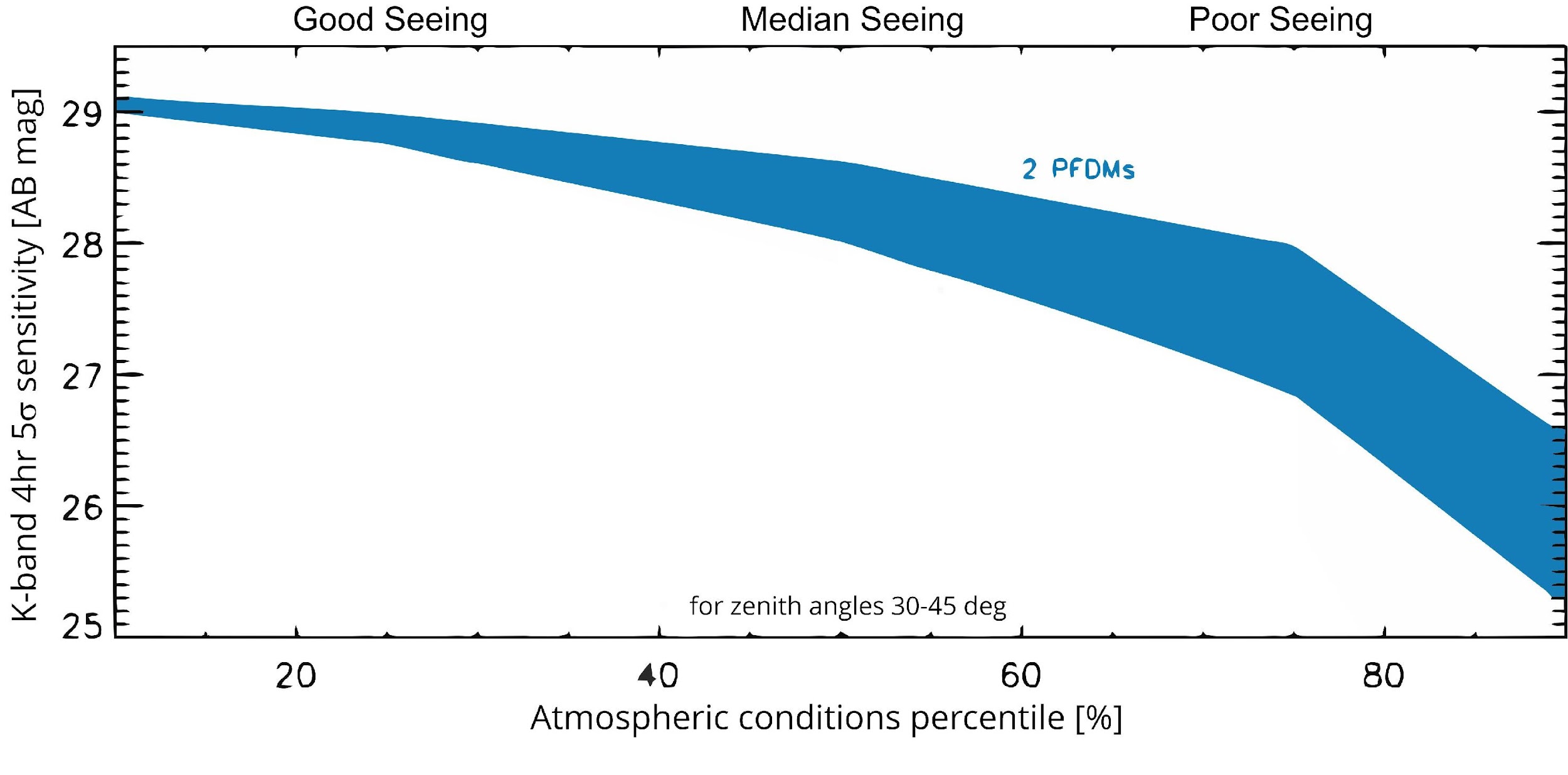}
    \caption{K-band sensitivity at 5$\sigma$ for a four-hour integration time for different atmospheric conditions.}
    \label{fig:KbandSensitivity}
\end{figure}
\begin{table}[ht]
\caption{Summary of the MORFEO error budget in nm RMS in the MICADO large FoV (r$\le$30 arcsec) for the 5 reference atmospheric profiles for coverage of half of the sky at the south galactic pole (the corresponding K band SR values are shown in Fig.~\ref{fig:comp_sky_cov_passata_baseline2020_2PFDM_budget180nm_K_SR}). Zenithal distance is 30deg. High order is the term associated with the modes sensed by the LGS WFSs while low order is the term associated with tip and tilt sensed by NGS WFSs (it depends on the magnitude and position of the available NGSs, that is the so-called sky coverage). Extra includes errors not accounted for in the end-to-end simulations (see text for details).}
\label{Tab:error}
\begin{center}
    \begin{small}
        \begin{tabular}{|l|c|c|c|c|c|}
            \hline
            \textbf{Error term} & \textbf{Q1} & \textbf{Q2} & \textbf{median} & \textbf{Q3} & \textbf{Q4}\\
            \hline
            High orders &  191 & 226 & 239 & 258 & 316 \\	
            \hline
            Low orders (*) & 137 & 161 & 169 & 185 & 231 \\
            \hline
            Extra & \multicolumn{5}{c|}{180} \\
            \hline
            Total & 296 & 331 & 343 & 365 & 431 \\
            \hline
            \multicolumn{6}{c}{\scriptsize (*) statistical value for half of the sky at the south galactic pole.}\\
        \end{tabular}
    \end{small}
\end{center}
\end{table}

In this section, we present the main results related to the AO performance of MORFEO.
The performance requirements of MORFEO are expressed in terms of K band SR for a zenith angle of 30deg: the average SR in the small science field of view (FoV, r$\le$10 arcsec) in the first quartile of atmospheric conditions must be greater than 50\%, and the average SR in the large science FoV (r$\le$30 arcsec) in the median atmospheric conditions over half the pointings in the sky must exceed 30\%.
The SRs are estimated through end-to-end simulations using PASSATA~\cite{doi:10.1117/12.2233963} and the sky coverage tool described in Ref. \cite{2022JATIS...8b1509P}.

The simulation parameters can be found in Ref. \cite{2020SPIE11448E..2SA}, Table 1, with a few exceptions: we updated the throughput considering 0.26 for the LGS path, 0.32 for the NGS path in the infrared and 0.22 in the visible, the LGS WFS configuration considering 68$\times$68 sub-aperture with a FoV of 16.1 arcsec and the NGS WFS configuration considering a pixel pitch of 14 mas.

The error budget is described in Tab. \ref{Tab:error}.
Part of the error budget, 180nm RMS (``Extra'' in the table), is not directly included in the end-to-end simulations, so it is added in the PSF calculation with random phase-screens with a $f^{-2}$ power spectral density.
This error includes terms such as non-common path aberration, residual aberrations in the telescope and in the post-focal relay optics due to design, manufacturing, alignment, thermo-elastic and optical effects of air in the optical path, aberration due to sodium layer variations, calibration errors, atmospheric chromaticity, and a contingency. 
In Fig. \ref{fig:SR_Qprof_e180nm} we report the average SR as a function of the off-axis angle for the 5 reference atmospheric profiles in case of a good NGS asterism: the K band SR in the small science FoV for the first atmospheric quartile (Q1, red line) is greater than 58\% (requirement $>$50\%). 
Performance as a function of the sky coverage at the south galactic pole is shown in Figs. 
\ref{fig:comp_sky_cov_passata_baseline2020_2PFDM_budget180nm_K_SR} and \ref{fig:comp_sky_cov_passata_baseline2020_2PFDM_budget180nm_JHK_fwhm}: the K band SR in the large science FoV for the median atmospheric condition (black line) is greater than 38\%.
Therefore, at the current stage of development, the system is in compliance with the performance requirements.

Note that these optical perfomance bring to transformational scientific merit function.
For example, in  Fig.~\ref{fig:KbandSensitivity},  we show the K-band point-source sensitivity at 5$\sigma$ for different atmospheric conditions, here computed using ELT telescope and sky emissivities and the MICADO detector characteristics (and the 4mas pixelscale configuration).
Additional information on the MORFEO + MICADO PSFs can be found in Ref.~\cite{2022SPIE12185E..5PA}.


\section{Conclusion}

We reported on progress of the MORFEO multi conjugate adaptive optics system,  which has reached the final design phase. 
We summarized its science goal, the design of its optics, wavefront sensors, deformable mirrors and real time computer, the development of calibration and test plans and we finally provided details on operation and performance and we showed that the system is in compliance with the performance requirements.

At this stage, system engineering activities~\cite{2022SPIE12187E..1OR} will complete the flow-down of requirements to all subsystems and subsequent verification, so that all design aspects are finalized and the next phase of Manufacturing, Assembly, Integration and Verification (MAIV) can begin.

Finally, note that despite the large amount of information reported above there are other AO related activities being carried out that have not been described in depth, such as 
the analysis of the error terms of the astrometric observations (for example we studied the tip-tilt anisoplanatism in MCAO systems and its impact on astrometric observations with next-generation telescopes in Refs. \cite{CarlaMNRAS,CarlaSPIE}) and research activities like the study on the so-called petalometer that is a sensor for differential segment pistons to deal with low wind effect\cite{2022A&A...665A.158P} and other related errors (currently we are looking for concepts like using LIFT in the LO WFSs \cite{2022SPIE12185E..56A} and CIAO-CIAO \cite{CarlaAO4ELT7}).




\printbibliography 

\end{document}